\documentclass{aa}         

\begin{document}

\thesaurus{12          
             (12.03.4; 
              12.12.1) 
           }

\title{Lema\^{i}tre-Tolman-Bondi model: fractality, bang time, and Hubble law}
\subtitle{I. Initial conditions and compatibility of density and
velocity laws}

\author{A. Gromov
        \inst{1}
   \and
        Yu. Baryshev
       \inst{2}
   \and
        D. Suson
        \inst{3}
   \and
        P. Teerikorpi
        \inst{4}
       }

\institute{
           Department of Computer Science
           St.-Petersburg State Technical University
           29 Politechnicheskaja str., 195251, St.-Petersburg, Russia.
           E-mail: gromov@natus.stud.pu.ru
      \and
           Astronomical Institute of St.-Petersburg University
           198904, St.-Petersburg, Russia.
           E-mail: yuba@astro.spbu.su
      \and
           Physics Department
           MSC 175
           Texas A$\&$MUniversity-Kingsville
           Kingsville, TX 78363, USA.
           E-mail: D-Suson@tamuk.edu
      \and
           Tuorla Observatory
           University of Turku
           FIN-21500 Piikki$\ddot {\rm o}$, Finland.
           E-mail: pekkatee@deneb.astro.utu.fi
         }

\offprints{P. Teerikorpi}
\mail{P. Teerikorpi}
\date{Received, 1999 / accepted, 1999}

\maketitle

\begin{abstract}

We start a systematic study of the Lema\^{i}tre-Tolman-Bondi
 (LTB) model as applied to
the large scale structure and its evolution.  Here we study
three possible initial conditions of the LTB models which are
asymptotically FRW at large scales: bang time, fractal density
(with fractal dimension $D=2$), and velocity law.  Any two of these
determine the third one. Fractal density and simultaneous bang time
provide a quantitative
estimate for the scale beyond which the deflection from the
linear Hubble law is small.
This border may be identified with the zero-velocity surface.
 For fractal density and linear Hubble law it is shown that the bang time
is necessarily non-simultaneous.

\end{abstract}

%

\section{Introduction}

Lema\^{i}tre-Tolman-Bondi (LTB) models are exact solutions of
Einstein's
equations for 1) spherical symmetry, 2) pressureless
matter (dust) and 3) motion with no particle layers intersecting. Originally
studied by Lema\^{i}tre (\cite{Lemaitre}), Tolman (\cite{Tolman}) and Bondi
(\cite{Bondi}), these models are the simplest generalization of the
Friedmann-Robertson-Walker (FRW) models with a non-zero density gradient.

At least three cosmological applications of LTB models have
been discussed. The first relates to the evolution
of inhomogeneities and peculiar velocities in an expanding universe.
Important results have been
derived under the assumption of ''unique bang time'' (Olson \& Silk
1979), when
every mass shell has been simultaneusly created.
So, Silk \& Wilson (\cite{Silk1}, \cite{Silk2}),
 Olson \& Silk (\cite{Olson1}), and
Olson \& Stricland (\cite{Olson2})
studied the formation of galactic clusters from small density and
velocity
perturbations (implicitly, non-simultaneous bang is used in
Silk \& Wilson (1979b)). It was shown that after a
long time the initial conditions are forgotten and a universal
density profile is formed.
The LTB model has been applied to the determination of the mass density
parameter $\Omega_o$ from the local peculiar
velocity field (Silk 1974), and of the
mass of the Virgo cluster (Hoffman et al. 1980; Tully \& Shaya 1984;
 Teerikorpi et al. \cite{Teerikorpi1}; Ekholm et al. 1999).

The second area of cosmological application was developed by Ellis
et al. (1985), Maartens (\cite{Maartens}) and Mustapha et al.
(\cite{Mustapha}). They use the FRW model for small (observable) scales
and assume the LTB model for large scales.

The third area is the modeling of fractal
matter distributions within general relativity
(Bonnor \cite{Bonnor}; Ribeiro \cite{Ribeiro1},
\cite{Ribeiro2}, \cite{Ribeiro3};
Humphreys et al. \cite{Humphreys2}; Matravers, \cite{Matravers}).
This has gained
impact from  redshift surveys revealing fractality with
the fractal dimension $D \approx 2$
in the
space distribution of galaxies up to distances of 100$h^{-1}$ Mpc
($h = H_0/100 {kms^{-1}Mpc^{-1}}$)
(see Baryshev et al.
\cite{Baryshev1}; Sylos Labini et al. \cite{Sylos}), and
confirming the scale invariant de Vaucouleurs
(\cite{deVaucouleurs}) law.  This
leads to a new application of the LTB models,
pointed out by Bonnor (1972),
where the fractal structure
is treated as spherically symmetrical inhomogeneities centered in
 every galaxy.

Baryshev et al. (\cite{Baryshev})
showed that the
linear perturbation theory for the growth of
density fluctuations leads to a non-linear Hubble
law if all matter is in the fractals.
Then the observed linear Hubble law (at scales $<$ 100$h^{-1}$
Mpc) requires a very low background (FRW) density, $\Omega_o <
10^{-3}$. The exact LTB calculations by Humphreys
et al. (\cite{Humphreys2})  and Matravers (\cite{Matravers}) gave the
same conclusions.
Another explanation for the linear Hubble law within
the fractal structure (the ``Hubble-de Vaucouleurs paradox'')
was proposed by Baryshev et al. (1998):
uniform dark matter with a very high density.
\footnote{Sandage et al. (\cite{Sandage1})
were the first to note the strange co-existence of
the linear Hubble law and {\it local} inhomogeneities.
Recently, Teerikorpi et al. (\cite{Teerikorpi2})
showed from Tully-Fisher distances
that the all-sky average number density decreases as predicted by a fractal
with dimension $\approx 2.2$, from $1$ to $100 h^{-1} Mpc$, where
(and beyond which) the Hubble law is seen.}
The
present paper shows a third way to make the linear
Hubble law, by abandoning the assumption of a unique bang time.

There are two ways to parameterize
LTB models. The first, introduced by Tolman and Bondi, is
called the $3+1$ approach.
The second uses
observational coordinates (Ellis et al. \cite{Ellis}).
As underlined by Matravers
(1998), the $3+1$
coordinate approach, adopted in the present paper, provides a physical
interpretation of the evolution of
the universe in co-moving coordinates.

Our series of papers is concerned with the 1st and 3rd
of the above areas and their theoretical basis. In this paper we
study initial conditions and compatibility of density and velocity laws. In
Paper II we
add the $\Lambda$-term to the LTB model,
and in Paper III address conceptual problems.

In Sect. \ref{review} we review the LTB models in co-moving
coordinates.
In Sect. \ref{sec 3} we study conditions for the existence of
the zero-velocity surface in the closed central part of the
LTB model.  We find  two criteria to check out if the zero-velocity
surface exists.
In Sect. \ref{sec 4} we show the velocity
deflection from the Hubble law within LTB models with unique bang time
and density law with fractal dimension $D \approx 2$, apply the
criterium
for the existence of the zero-velocity surface, and show
how its location
depends on initial conditions and cosmological density
$\Omega_0$.
We also calculate the bang time function which reproduces the
linear Hubble law at all scales within the inhomogeneities described
 by LTB models at the present cosmic epoch.

\section{The LTB models in co-moving coordinates}

\label{review}

The LTB models are the simplest exact, nonlinear,
inhomogeneous, nonstationary, isotropic dust models in general relativity.
In
co-moving (Lagrangian) and synchronous coordinates $r$, $t$ the metric is:
\begin{equation}
\label{metric}
{d}s^2(r,t)=c^2\,{d}t^2-{e}^{\lambda (r,t)}
\,{d}
r^2-R^2(r,t){d}\Omega^2
\end{equation}
and  the energy-momentum tensor has the form:
$T_k^i= diag(c^2\, \rho(r,t),0,0,0)$.
Here $c$ is the speed of light, ${d}\Omega^2 = {d}\theta^2
+ \sin
\theta\,{d}\phi^2$, and
$R(r,t)$ is an Euler coordinate. Such metrics and
energy-momentum tensor, together with Einstein's equations,
produce a set
of inhomogeneous cosmological models, generally with time- and
space-dependent curvature, which are described by the following
system (Tolman 1934):
\begin{eqnarray}
\frac{8\,\pi\,G}{c^4}\,T_0^0 = -\frac{{e}^{-\lambda}}{R^2}\,
\left(
2\,R\,R^{\prime\prime} + R^{\prime\,2} - R\,R^{\prime}\,\lambda^{\prime}
\right) + \nonumber\\
\frac{1}{R^2}\,\left(
R\,\dot R\, \dot \lambda + \dot R^2 + 1
\right),
\label{E equations 0 0}
\end{eqnarray}
\begin{equation}
\frac{8\,\pi\,G}{c^4}\,T_1^1 = \frac{1}{R^2} \,\left(
-{e}^{-\lambda}\,R^{\prime\,2}\,
+
2\,R\,\ddot R +  \dot R^{2} + 1 \right) = 0,
\label{E equations 1 1}
\end{equation}
\begin{eqnarray}
\frac{8\,\pi\,G}{c^4}\,T_2^2 = -\frac{{e}^{-\lambda}}{R}\,
\left(
R^{\prime\prime} - \frac{R^{\prime}\,\lambda^{\prime}}{2}
\right) + \nonumber\\
\frac{\dot R\,\dot\lambda}{2\,R} +
\frac{\ddot \lambda}{2} + \frac{\dot\lambda^2}{4} +
\frac{\ddot R}{R} = 0,
\label{E equations 2 2}
\end{eqnarray}
\begin{equation}
T^1_0 = \frac{{\rm e}^{-\lambda}}{R}\,
\left( 2\,\dot R^{\prime} - \dot \lambda\,R^{\prime} \right) = 0,
\label{E equations 1 0}
\end{equation}
\begin{equation}
T^2_2 = T^3_3,
\label{E equations 2 2 3 3}
\end{equation}
where ${}^{\prime }=\frac{\partial}{\partial r}$ and
$\dot{}=\frac{\partial}{\partial (c\,t)}$.
Einstein's equation Eq.(\ref{E equations 1 0})
defines the function $\lambda (r,t)$:
\begin{equation}
\label{metric 1}
{e}^{\lambda (r,t)}=R^{\prime \,2}(r,t)/f^2(r).
\end{equation}
$f(r)$ is one of the
undetermined functions of the models.
Here ${}^{\prime }=\frac{\partial}{\partial r}$ and
$\dot{}=\frac{\partial}{\partial (c\,t)}$.
Via Eq.(\ref{metric 1}), the $T^1_1$-component reduces to
the equation of motion
\begin{equation}
\label{eq-m}
2\,\ddot R(r,t)R(r,t)+\dot R^2(r,t)+1-f^2(r)=0.
\end{equation}
The first and second
integrals of Eq.(\ref{eq-m}) are:
\begin{equation}
\label{TB-klass}
\dot R^2(r,t)=\,f^2(r)-1 + \frac{\,F(r)}{2R(r,t)},
\end{equation}
\begin{equation}
\label{T:15:11}
\pm t+t_R(r)=\int\limits_{R(r,t_0)}^{R(r,t)} \frac{{d}\tilde R}{\sqrt{c^2\,(f^2(r)-1)+%
\frac{F(r)}{2\tilde R}}},
\end{equation}
where $F(r)$ and  $t_R(r)$ are the second and third undetermined functions
of the models. $t_R(r)t_R(r)$ is the
bang time $t_R(r)$.
Equation (\ref{E equations 0 0}) gives for the density:
\begin{equation}
\label{T:15:1}
\frac{8\,\pi\,G}{c^4}\,T_0^0(r,t) =
\frac{{d}F(r)}{{d}r}\,
\frac{1}{2\,R^2(r,t)\,\frac{\partial R(r,t)}{\partial r}},
\end{equation}
Eq.(\ref
{T:15:1}) shows that the models become singular by two different causes,
 defined by $R(r,\tau )=0$ (which corresponds to the bang time $t_R(r)$) and
$R^{\prime }(r,\tau )=0$ (which corresponds to the layer intersection
time $t_{R^{\prime}}(r)$).
The bang time
(Silk \& Wilson 1979a)
and layer intersection time function (for flat LTB model see Gromov
\cite{Gromov2}),
respectively, correspond to these singularities.
The
gravitating mass $M_\mathrm{grav}$ and the invariant mass $M_\mathrm{inv}$
of the dust are defined by the energy-momentum tensor:
\begin{equation}
\label{f phys sense 1}
M_{grav}(R(r,t))=\frac{4\pi}{c^2} \,\int\limits_0^{R(r,t)} T_0^0(y)\,
y^2\,{d}y = M_{grav}(r)
\end{equation}
and
\begin{eqnarray}
\label{f phys sense 2}
M_{inv}(r)=\frac{4\,\pi}{c^2} \int\limits_0^rT_0^0 (x,t)\,\sqrt{%
-g(x,t)}{d}x,
\end{eqnarray}
where $\sqrt{-g} = \frac{R^{\prime}\,R^2}{f}$.
Substituting $\rho $ from the definition of $T_i^k$ and
Eq.(\ref{T:15:1}) into Eq.(\ref{f phys sense 1}) and
Eq.(\ref{f phys sense 2}) we obtain
$M_{grav}(r)=(c^2/4\,G)\,F(r)$, and F(0)=0.
So,
\begin{equation}
\label{f phys sense 4}
M_{inv}(r)=\int\limits_0^r\frac{{d}M_{{grav}(x)}}{f(x)%
},\qquad f(r)=\frac{M_{grav}^{\prime }}{M_{inv}^{\prime }}.
\end{equation}
One of the interpretations of the function $f(r)$,
given by Bondi (1947), relates it
to the curvature and components of the Einstein tensor:
$$
K^{1}_{1} = 2\,\frac{f}{R}\,\frac{{d}f}{{d}R}\qquad
K^{2}_{2} = K^{3}_{3} = \frac{f^2 - 1}{R^2} + \frac{f}{R}\,\frac{
{d}f}{{d}R},\qquad
$$
$$
{\bf K}=\frac 2{R^2}\,\frac {{d}}{{d} R}\left( R\,\left( f^2-1
\right)
\right).
\label{curva}
$$
The space curvature is equal to zero if and only if $f=1$.

The LTB models are defined up to some transformation of the co-moving
coordinate $\psi: r \to \tilde r$, which decreases the number of undetermined
functions from 3 to 2 (Just \cite{Just1}; Just \& Kraus \cite{Just2}).
See also Hellaby \& Lake (\cite{hellaby85}) and Hellaby (\cite{hellaby87})
for interesting examples.
These two
functions should be chosen from the set:
\begin{eqnarray}
t_R, \quad t_{R^{\prime}}, \quad \rho_0, \quad f, \quad R_0, \quad  \dot R_0.
\label{set}
\end{eqnarray}
The transformation is not unique and may be chosen according to the
character of the problem (see Gromov (\cite{Gromov1}).

The transformation $\psi$ is time independent, so it can be used to fix one
of the functions from the set Eq.(\ref{set}). In this case one speaks about a
parametrization of the LTB model. We give two examples. The first is
often used:

\begin{equation}
\label{transformation 1}
R(r,0)=r.
\end{equation}
This implies that $F(r)$ is defined by the initial density
profile $\rho (r,0)$. It was fully studied by Liu (\cite{Liu1},
\cite{Liu2},\cite{Liu3}), and used e.g. by Ribeiro (1992a, 1992b, 1993)
and Gon\c{c}alves \& Moss (\cite{Goncalves}).  An alternative
transformation is:
\begin{equation}
\label{transformation 2}
M_{inv}(r)\,G/c^2=r.
\end{equation}
It was first used by Eardley (\cite{Eardley}) and studied by Gromov
(\cite{Gromov1}, \cite{Gromov2}).

The bang time $t_R$ is used as one of the initial conditions
by Silk \& Wilson (1979a), Olson \& Silk (1979),
Olson \& Stricland (1990).

\section{Zero-velocity surface in the LTB model}
\label{sec 3}

We use two different sets of initial
conditions: \newline
A) bang time $t_R(r)$ and initial density profile $\rho _0(r)$, and \newline
B) initial density $\rho _0(r)$ and velocity $\dot R_0(r)$ profile.

They are interconnected as may be seen from a simple analogy of
an apple dropping from
an apple tree. If the initial position and velocity, and an equation of
motion are given, we can calculate the time it will reach the
ground. For the LTB models it is the same. If we start with $%
t_R(r)$ and $\rho _0(R)$ as initial conditions, we can calculate the
velocity profile $\dot R(r)$ for $t=0$: since a particle must
reach the centre by the time $t_R(r)$, it must have a predefined
velocity at $t=0$. Since the density profile is also given, the
gravitational potential becomes fixed by the same initial conditions.
As from each two functions the third may be derived, all three cannot
be selected freely.

\subsection{Dimensionless equations}

\label{sec3-1}

Now we restate the models in terms of dimensionless
quantities. We use the following characteristic values:
$$
l_0 = c\,t_0, \qquad t_0 = \frac{1}{H_0},  \qquad
\Omega_0 = \frac{\rho(\infty)}{\rho_{cr}},
$$
$$
8\,\pi\,G\,\rho_{cr} = 3\,H^2_0\qquad
M_0 = \frac{4\,\pi}{3}\,\rho(\infty)\,l^3_0,
\label{d l v 1}
$$
and dimensionless variables:
$$
\xi = \frac{R}{l_0}, \qquad \eta = \frac{r}{l_0}, \qquad
\tau = \frac{t}{t_0}, \qquad \delta(\xi) = \frac{\rho(R)}{\rho(\infty)},
$$
$$
\mu(\eta) = \frac{M_{grav}}{M_0} = 3\,\int\limits_{0}^{\eta}
\delta(x)\,x^2\,
{d}x,
\label{d l v 2}
$$
where $l_0$ and $t_0$ are the characteristic length and time,
$\Omega_0$ is the density parameter of the FRW background,
$H_0$ is the value of the Hubble parameter at the moment of
the initial conditions, $\rho_0(r)$ is the dust density, $\rho _{cr}$ is
the critical density, $M_0$ is the characteristic mass, $\xi $ and $\eta$
are dimensionless Euler and Lagrangian coordinates,
$\delta (\eta )$ is the dimensionless
density, and $\mu (\eta )$ is the dimensionless mass of the dust. In terms
of these quantities, the bang time is written as $\tau _\xi (\eta)$. The
index $\xi $ reminds us that the bang time is the time required for
the particle to come from its initial position $\xi_0$ to $\xi = 0$.

In this section we show how one can use the first and second integrals,
Eq.(\ref{TB-klass}) and Eq.(\ref{T:15:11}), to calculate the zero-velocity
surface for LTB models if we relax the often used assumption of
unique bang time.

We use an effective dimensionless gravitating mass $\mu ^{*}$
\begin{equation}
\label{m eff}
\mu ^{*}=\Omega _0\,\mu = 2 M_{{grav}}G/c^2l_0.
\end{equation}
In terms of dimensionless variables the first integral of the equation of
motion Eq.(\ref{TB-klass}) becomes
\begin{equation}
\label{TB klass d}
\dot \xi ^2(\eta,\tau )=f^2(\eta)-1+\frac{\mu ^{*}}{%
\xi (\eta ,\tau )}.
\end{equation}
The form of the second integral Eq.(\ref{T:15:11}) depends on the
sign of $f^2(\eta)-1$.
 For $f^2(\eta)-1<0$ (closed models):
\begin{eqnarray}
\pm\tau + \tau_{\xi}(\eta) = \nonumber\\
\frac{\mu^{*}(\eta)^{-1/2}}{A^{3/2}}\,
\left(
{arcsin}\sqrt{A} -
\sqrt{A} \,
\sqrt{1 - A}
\right);
\label{closed}
\end{eqnarray}
for $f^2(\eta )-1=0$ (flat models):
\begin{eqnarray}
\xi^{3/2} = \xi^{3/2}_0 \pm \frac{3}{2}\,\tau\,\sqrt{\mu^{*}};
\label{flat}
\end{eqnarray}
and for $f^2(\eta) - 1 > 0$ (open models):
\begin{eqnarray}
\pm\tau + \tau_{\xi}(\eta) = \nonumber\\
\frac{\mu^{*}(\eta)^{-1/2}}{(-A)^{3/2}}\,
\left(-{arcsinh}\sqrt{-A} + \sqrt{-A} \sqrt{1 - A}\right)
\label{open}
\end{eqnarray}
In these formulas:
\begin{equation}
A = \frac{1 - f^2(\xi)}{\mu^{*}(\eta)}\,\xi(\eta,\tau).
\end{equation}

\subsection{The closed and open models with arbitrary bang time}

\label{sec3-2}

In this and the remaining sections we concentrate on initial
conditions. The co-moving coordinate is chosen as $\xi_0$.
By solving Eq.(\ref{TB klass d}) for $1-f^2$ and substituting this
into Eq.(\ref{closed}), the expression for bang time $\tau _\xi $ of
the closed and open models may be rewritten in the form (taking into
account that for a fixed moment, $\xi$ may be used as a co-moving coordinate):
\begin{eqnarray}
\label{closed 1}
\tau _\xi=\sqrt{\frac{\xi ^3}{\mu ^{*}}}\,\Psi, \qquad
B=\frac{\xi_0 \dot \xi_0 ^2}{\mu ^{*}} \geq 0,
\end{eqnarray}
and for closed models $0 \leq B < 1$
\begin{eqnarray}
\label{closed 2}
\Psi (B)\equiv \Psi ^{cl}(B)=\frac{{arcsin}\sqrt{1-B}-%
\sqrt{1-B}\,\sqrt{B}}{(1-B)^{3/2}},
\end{eqnarray}
while for open models $B > 1$
\begin{eqnarray}
\label{closed 3}
\Psi (B)\equiv \Psi ^{op}(B)=\frac{-{arcsinh}\sqrt{B-1}+%
\sqrt{B}\,\sqrt{B-1}}{(B-1)^{3/2}}.
\end{eqnarray}
Eq.(\ref{closed 1}) allows us to rewrite equation Eq.(\ref{TB
klass d}) as
\begin{equation}
\label{i 2 new}
B=(f^2-1)\,\frac {\xi_0}{\mu ^{*}}+1.
\end{equation}
It follows from Eq.(\ref{closed 1}) that $B = 0$
corresponds to the following set of initial conditions: if $\xi_0 = 0$,
$ grad \delta(\xi_0 = 0) = 0$, when
\begin{equation}
\lim_{\xi_0 \to 0}B \sim \lim_{\xi_0 \to 0}\left(\dot
\xi_0/\xi_0\right)^2 \geq 0;
\label{set i c 1}
\end{equation}
whereas in the case of $\xi_0 \ne 0$, $B = 0$ implies
$\dot \xi_0 = 0$.
The velocity of the particle is equal to zero at
the boundary $\xi_{ZV}$, see Figs. 1-3.
For both cases $f^2 = 1 - \mu^{*} / \xi_0 \geq 0$ for $B = 0$,
which implies the inequality
$\xi_0 \geq \mu^{*}$ for $B = 0$.
Note that this restricts the kind of particular TB model in
which  the nonequality may be satisfied: because $f^2 \geq 0$, it follows
that $f^2 - 1 < 0$. So, $B = 0$ may be satisfied
only in the closed model.

The limit $B\to 1$ corresponds to $f\to 1$, so that open and closed
models both have a common limit which coincides with the flat model:
$\Psi ^{fl}(B)=\frac 23$.

Olson \& Silk (1979) defined the boundary between open and
closed TB models with a unique bang time as
a place where $f=1$.
We also postulate a set of
initial conditions (e.g. fractal density and
Hubble law), which produce the following sequence of particular models: a
closed domain of the model which has a position around a centre (''core'')
 and open model
farther out from the centre (''shell'').
The two domains are separated by the flat one located on the surface where
 $%
f^2=1$.
For the closed "core" $0 \leq B < 1$,
so, from Eq.(\ref{closed 2}) and Eq.(\ref{i 2 new}) it follows that
\begin{equation}
\label{B bound 1}
\frac 23<\Psi^{cl}(B)\leq {arcsin}(1)\approx 1.57,
\end{equation}
where $\frac 23$ corresponds to the boundary of the closed
model, the flat model,
(this boundary we denote by $\xi _{fl}$) and arcsin(1)
corresponds to the zero-velocity surface $\xi _{ZV}$.

For arbitrary bang time, the zero-velocity surface is defined by
the solution of Eq. (\ref{def of TB domain 1}):
\begin{equation}
\tau_{\xi} = \sqrt{\frac{\xi_0^3}{\mu^{*}}}\,\arcsin(1).
\label{def of TB domain 1}
\end{equation}
The solution of Eq.(\ref{def of TB domain 1}) may be real or complex
depending on initial conditions, i.e. bang time and density profile.
A complex solution means that the zero-velocity surface lies in
the centre of symmetry. If the solution is real (and positive) then
the zero-velocity surface is found at a finite distance from the centre
and separates the collapsing region of the closed part of the model
from the expanding region.
 In the domain $0 \leq \xi_0 < \xi_{ZV}$ we can
introduce  some other LTB model corresponding to the initial density profile,
also closed. Humphreys et al. (\cite{Humphreys1}) first
demonstrated how to construct the LTB model for that domain.

The above approach utilizes the \emph{coordinate criterium} for the
existence of a central collapsing domain. Using
Eq.(\ref{closed 1}), we can also define a second form for this criterium, the
\emph{mass criterium}. From Eq.(\ref{closed 1}) it follows that the two
limits of
function $\Psi (B)$ correspond to two characteristic masses. The mass $\mu
_{ZV}$,
\begin{equation}
\label{uuu-tb}
\mu _{ZV}(\xi_0)=\left(1.57/\tau _\xi\right)
^2\,\xi_0 ^3,
\end{equation}
corresponds to the low limit of radial Euler coordinate $\xi _{ZV}$.
This designates the beginning of the domain $\xi_0 > \xi_{ZV}$
 from which all particles can collapse at time
$\tau _\xi$. Similarly, the characteristic mass corresponding to the
flat model (or to the upper boundary of the closed model, which is the same
thing) has the form
\begin{equation}
\label{uuu-fl}
\mu _{fl}(\xi_0 )=\left(2/3\,\tau _\xi\right)
^2\,\xi_0 ^3.
\end{equation}
This criterium can be stated as follows: if the graph of the mass,
corresponding to a given initial density profile, intersects the graph of $%
\mu _{ZV}$, then $\xi _{ZV}>0$.

\subsection{The flat LTB model}

\label{sec3-3}

We now turn to the simplest initial condition, $f=1$. If $\tau _\xi
={const}$, the flat LTB model reduces to the flat FRW model. As
shown by Gromov (1997), for the flat LTB model the bang time
may be given in the form:
\begin{equation}
\label{t1}
\tau _\xi (\mu )=\sqrt{\frac 1{\mu ^{*}}\,\int\limits_0^{\mu}%
\frac{{d}y}{\rho _0(y)}},
\end{equation}
which immediately implies that $\rho _0(\xi_0 )={const}$ for
simultaneous
bang time. In any other case, $\rho _0(\xi_0 )\ne {const}$ and the bang
time is not constant. As was shown above, $\xi _{ZV}$ may not
 be
equal to zero (and the bang is not simultaneous)
if and only if the LTB model is closed, so the domain of
definition of the flat model is the whole region $\xi_0 \geq 0$.

\section{LTB models for a fractal density distribution}

\label{sec 4}

In this section we study the LTB models with initial conditions
given by the fractal density profile and Hubble law.

\subsection{Fractal density distribution and the Hubble law: Hubble-de
Vaucouleurs paradox}

\label{4_1}

Two fundamental empirical laws have been established from
extragalactic data. First, there is the power law density-distance relation
(de Vaucouleurs law) which corresponds to fractal struture
(Mandelbrot \cite{Mandelbrot})
with
fractal dimension $D\approx 2$ up to the depth of available catalogs
$\approx$ 100 $h^{-1}$ Mpc. (Sylos Labini et al. 1998).
Second, Cepheids,
TF-distance indicator and Type Ia supernovae
confirm the linearity of Hubble's redshift-distance law within the same
distances where the fractality exists.

Baryshev et al. (1998) emphasized that the linear
redshift-distance relation inside the fractal (inhomogeneous)
matter distribution creates the so-called Hubble-de Vaucouleurs (HdeV)
paradox. It means that the interpretation of the Hubble law within FRW
cosmological models as a consequence of a homogeneous galaxy
distribution disagrees with new data on the galaxy
distribution for a scale interval from 1 to 100 Mpc. We emphasize that the
paradox exist for small distance
scales (up to 100 Mpc), i.e. for redshifts less then 0.03. Hence the arguments
of Abdalla et al (\cite{Abdalla}) on essential relativistic corrections
do not explain the paradox.

Two solutions of the HdeV paradox are previously known. The first one
(Baryshev et al. 1998) is based on uniform
\emph{dark} matter starting just from the halos of galaxies, in which
case the standard FRW model works. But then the fractal
distribution of \emph{luminous} galaxies can appear only from
 special
initial perturbations of FRW.
\footnote{But, de Vega et al. (\cite{deVega})
 and de Vega \& Sanchez (1999) showed
that Newtonian self-gravitating
N-body systems have a quasi-equilibrium fractal state
with a dimension of $\approx 2$.}

The second solution is a very low value for the
global average density (Baryshev et al. 1998;
Humphreys et al. 1998b).
However, if the upper
cut-off scale of the fractal structure is large, the low density
contradicts the estimated density of the
baryonic luminous and dark matter.

\subsection{On the applicability of the LTB model to fractals}

\label{4_2}

The LTB model has proved useful for understanding the kinematics
of galaxies around \emph{individual} mass concentrations. For
example, Teerikorpi et al. (1992) could put in evidence the
expected behaviour in the Virgo supercluster: 1) Hubble law at
large distances, 2) retardation at smaller distances, 3) zero-velocity
surface, and 4) collapsing galaxies at still smaller distances.

Bonner (1972) was the first to apply the LTB model to the hierarchical
 cosmology.
He used de Vaucouleurs's
density law $\rho \sim d^{-\gamma}$ with $\gamma =1.7$. Ribeiro
(1992a, 1992b, 1993) has developed a numerical
approach to solving LTB equation for fractal galaxy distribution.
Humpreys et al. (1998b) gave a
relation between number counts and redshifts for LTB models with
large scale FRW behaviour.

However, the application of LTB models to a fractal distribution
leads to a conceptual problem, because the original
LTB formulation contained a central
point of the universe, around which the density distribution is isotropic.
In a fractal distribution (Mandelbrot \cite{Mandelbrot}) there is no
unique centre, but every object of the structure may be treated as a local
centre which accommodates the LTB centre.  Every structure point is surrounded
by a spherically symmetric (in average) matter distribution.

In this sense, the application of the LTB model to fractals means that there
is an infinity of LTB exemplars with centres on every structure point.  Their
initial conditions are slightly different, because for any fixed scale the
average density is approximately constant.  For different scales the density
is a power law. This excludes geocentrism and makes
possible the use of LTB models as an exact general relativistic
cosmological model where expansion of space becomes scale
dependent.
\footnote{A similar property exists
for Friedmann models where the rate of space expansion within distance
$d$ from a fixed galaxy is determined by the total mass of the sphere around
this galaxy. For LTB
fractal models the space expansion at distance $d$ from a fixed point of the
fractal structure is also determined by the
average mass of the sphere around this point (Paper III).}

\subsection{Simultaneous bang time}

\label{4_3}

We showed in Sect.{3} that a unique bang time and
constant density imply open FRW models. Here, with nonlinear LTB
models, we study density perturbations with arbitrary amplitude
on the FRW background
and show how the initial fractal density changes
the models.

Humphreys et al. (\cite{Humphreys2}) used a density profile which
needed junction conditions for
densities corresponding to different scales. We
consider the analytical case of a smooth density profile.
The fractal density on the FRW background and simultaneous
bang are the initial conditions of our LTB models:
\begin{equation}
\label{i cc 1b}
\delta (\xi ,0)=\frac A{\epsilon +\xi_0 }+1,
\end{equation}
\begin{eqnarray}
\tau_{\xi}(\xi_0) = \tau_{\xi}(\infty) = \tau_{FRW} = {const.}
= \nonumber\\
\frac{1}{(1 - \Omega_0)^{3/2}}\,
\left(
\sqrt{\frac{1 - \Omega_0}{\Omega_0}}\,\sqrt{\frac{1}{\Omega_0}}
-
\arcsin \sqrt{\frac{1 - \Omega_0}{\Omega_0}}\right)
\label{i cc 1a}
\end{eqnarray}
Here $\epsilon =R_{galaxy}/l_0 \sim 10\,{Kpc}/
5\cdot 10^6\,{Kpc}%
=2\cdot 10^{-6}$.
Above the galactic scale, Eq.(\ref{i cc 1b}) describes the fractal
density law with $D=2$.
The density contrast
($\rho _{galaxy}/\rho _{FRW})$ is $\delta (\xi =0)
\sim 10^{-24}\,
{g/cm^3}/
10^{-29}\,{g/cm^3}=10^5$. So, $A\sim 0.2$. For our
calculations
we use $A=0.002,\,0.02,\,0.2,\,2$, which imply the amplitude of the density
$\delta(0) = 10^{3},\,10^{4},\,10^{5},\,10^{6}$, and we use $\Omega_0 =
0.001,\,0.01,\,0.1,\,0.99$, which imply $\tau_{FRW} =
0.997,\,0.98,\,0.898,\,0.688$.

The properties of the LTB models with
initial conditions Eq.(\ref{i cc 1b}), Eq.(\ref{i cc 1a}) were studied
in Sect. 3.
Now we apply
the results of Sect. 3.2. to the initial
conditions with given parameters.
For the chosen values of $A$ and $\Omega_0$ LTB
models have a closed "core" and open "shell". But only for $\Omega = 0.001$
and $A = 0.002$ does Eq.(\ref{def of TB domain 1}) have a complex
solution (Fig.1).
\begin{figure}
\caption{
The solution of Eq.(\ref{def of TB domain 1})    with
unique bang time and initial conditions. $\xi_{{ZV}}$, the
solution of Eq.(\ref{def of TB domain 1}), depends on the initial
conditions (\ref{i cc 1b}) and (\ref{i cc 1a}).
The upper curve is the function
$\log{(\tau_{{FRW}}(\xi_0^3/\mu^*)^{1/2})}$
for $\Omega_0 = 0.001$ and $A = 0.02$ (see
Table 3), while the lower curve corresponds to $\Omega_0 = 0.001$ and $A =
0.002$ (Table 4).
The upper line is $\max {\rm log(\Psi^{cl})} = {\rm log(arcsin(1))} = 0.196$.
The lower line is $\min {\rm log(\Psi^{cl})} = {\rm log}(\frac{2}{3})
 = -0.176$.
The galaxy scale $\epsilon$ is $2 \cdot 10^{-6}$.
The curve intersects the upper line at $\xi_{{TB}}$.
It is seen that $\Omega_0 = 0.001$ and $A = 0.02$ produce the
intersection at a scale $>$ galaxy scale, which corresponds to real
(and positive) solution of Eq.(\ref{def of TB domain 1}),
but $\Omega_0 = 0.001$ and $A = 0.002$ do not produce it. In
the last case the solution of Eq.(\ref{def of TB domain 1}) is
imaginary.
}
\end{figure}
This means that only these parameters
produce the TB model with fractal density and simultaneous bang time with
$\xi_{ZV} \geq 0$ (see Table 4).
For all other cases of the adopted parameter values,
Eq.(\ref{def of TB domain 1})
 has a real solution and $\xi_{ZV} > 0$
 (see Tables 1 - 4).

Figs.2 and 3 illustrate the coordinate
and mass criteria for the
existence of $\xi_{{ZV}} > 0$  for $A = 0.02$ and $\Omega = 0.01$.

\begin{figure}
\caption{
The coordinate criterium of existence of
$\xi_{{ZV}}$ for simultaneous bang time
$\tau_{{FRW}} = 0.98$,
$\Omega_0 = 0.01$ and $A = 0.02$ (see Table 3).
The model is defined for $\xi_0 \geq \xi_{{ZV}}$. At the boundary
$\xi_{{ZV}}$
velocity $\dot \xi(\xi_{{ZV}}) = 0$, see Fig. 4.
}
\end{figure}
\begin{figure}
\caption{
The mass criterium of existence of
$\xi_{{ZV}} > 0$ for simultaneous bang time
$\tau_{{FRW}} = 0.98$,
$\Omega_0 = 0.01$ and $A = 0.02$ (see Table 3).
The upper of two parallel
lines corresponds to $\mu_{{TB}}$ and the lower line corresponds to
$\mu_{{fl}}$, see Eq.(\ref{uuu-tb}) and Eq.(\ref{uuu-fl}).
The model is defined for $\xi_0 \geq \xi_{{ZV}}$. At the surface
$\xi_{{ZV}}$
velocity $\dot \xi(\xi_{{ZV}}) = 0$, see Fig. 4.
}
\end{figure}
Our solution depends on $\Omega _0$: $\mu ^{*}\sim \Omega _0$ and
$\tau_{{FRW}}=\tau _{{FRW}}(\Omega _0)$.
Tables 1 - 4 show
characteristic values of $\xi _{{ZV}}$ and $\xi _{{fl}}$ for
different $A$ and $%
\Omega _0$. Here $l_0=c/H_0=5000 {Mpc}$, $H_0 = 60 {km^{-1}
s^{-1}Mpc}$.

Finally, we calculate the velocity $\dot \xi (\xi )$,
produced by the initial conditions Eq.(\ref{i cc 1b}) and Eq.(\ref{i cc 1a}).
For
both domains of the model, closed and open,
\begin{equation}
\label{vel}
\dot \xi_0 =\sqrt{(\mu ^{*}/\xi_0)\,B},
\end{equation}
where $B$ is the solution of the equation (27). Fig. 4 shows the
resulting non-linear velocity-distance relations down to the
zero-velocity radius, which in a natural manner marks the distance
close to which the deflection from the Hubble law becomes very
significant.

Tables 1-4 and Fig. 4 confirm the previous conclusion by
Baryshev et al. (1998) that the observed linear Hubble law
is compatible with such a fractal density only if the
FRW density parameter $\Omega_0$ is small.  For instance,
if $\Omega_0 = 0.99$, then the zero-velocity radius $R_{{ZV}}$
ranges 3.4 - 3400 Mpc for the range of the density contrast
$A = 0.002 - 2$, and is 344 Mpc for the ``preferred'' value
of $A=0.2$.  With a very small value of $\Omega_0$, 0.001,
$R_{{ZV}}$ appears around 0.6 Mpc, which is an intergroup scale,
while a good linear Hubble flow is reached around 6 Mpc.

\subsection{Nonsimultaneous bang time}

\label{4_4}

Here we study still another solution of the HdeV paradox within the
LTB model:
a non-unique bang time may produce
the linear Hubble law within a fractal structure having an
upper cut-off.

We take the initial conditions from
the observations in the domain from 1 Mpc to 100 Mpc, i.e. the fractal
density and the Hubble law:
\begin{eqnarray}
\delta (\xi ,0)=\frac A{\epsilon +\xi_0 }+1, \quad
\dot\xi = \xi.
\label{i 15}
\end{eqnarray}
The bang time in this situation is calculated by two formulas, corresponding
to the closed and open domains:
\begin{eqnarray}
\tau^{cl}_{\xi}(\xi_0) =
\frac{1}{(1 - B)^{3/2}}\,
\left(
-\sqrt{B}\,\sqrt{1 - B} \right. \nonumber\\
\left. + \arcsin \sqrt{1 - B}
\right),
\label{i c 1a}
\end{eqnarray}
\begin{eqnarray}
\tau^{op}_{\xi}(\xi_0) =
\frac{1}{(B - 1)^{3/2}}\,
\left(
\sqrt{B}\,\sqrt{B - 1} - \right. \nonumber\\
\left. {arcsinh} \sqrt{B - 1}
\right),
\label{i c 1b}
\end{eqnarray}
where from Eq.(\ref{open}) and Eq.(\ref{i 15}),
$B = \xi_0^3/\mu^{*}(\xi_0)$.
For $A=0.02$ and $\Omega _0=0.01$
the bang time $\tau _\xi (\xi_0 )$ is shown in Fig.5.
\begin{figure}
\caption{
Velocity produced by initial conditions Eq.(\ref{i cc 1b}) and Eq.(\ref{i cc 1a})
(simultaneous (FRW) bang time and given density profile)
for $\Omega_0 = 0.001, 0.01,0.1$ (from left to right) and $A = 0.02$.
 Different $\Omega_0$
produce different $\xi_{{ZV}}$, in which $\dot\xi(\xi_{{ZV}})
= 0$. Recall that $\log{\xi} = -1$ corresponds to 500 Mpc and the size
of the Local Group (1 Mpc) is encountered at $\log{\xi_0} = -3.7$.}
\end{figure}

\begin{figure}
\caption{
The non-simultaneous bang time
$\tau_{\xi}(\xi_0)$ produced by initial conditions (\ref{i 15}) with
$\Omega_0 = 0.01$, $A = 0.02$.
At infinity the time of collapse $\tau_{\xi}(\xi_0) \to \tau_{{FRW}}
 = 0.98$, indicated by the horizontal line.}
\end{figure}

\section{Discussion and conclusions}

\label{5}

We have studied the LTB model solution for two  pairs
of initial conditions related to observations: (bang time function,
 density profile) and
(density profile, velocity function) which both permit one to
treat the LTB problem as the Cauchy problem.  The second pair
may be obtained from observations at the present
epoch, while from the first pair, with bang time function assumed,
one may predict the present-day velocity function.  A
discussion of this mathematical side is given by Gromov (\cite{Gromov4}).
We show that for most parameter values of the model, there is the zero-velocity
surface.

Our study is much guided by Bondi's (1947)
idea when he
said ``The assumption of spherical symmetry supplies us
with a model which lies between the completely homogeneous models of
cosmology and the actual universe with its irregularities.''

As the ``actual universe with its irregularities'' has turned out to be
fractal, at least in its luminous matter and in scales up to 100 Mpc or more,
we follow some earlier works in representing fractality with spherically
symmetrical systems of different scales.  The interesting conceptual
difficulties will be treated in Paper III.  Here we have assumed that the
fractal representation is adequate and complement our previous work
(Baryshev et al. \cite{Baryshev}) on the Hubble law within fractals.
The qualitative conclusions of that paper, based on the linear regime,
are confirmed by our exact LTB solutions.  In particular,
a fractal distribution of matter with $D = 2$, smoothly going over
to the FRW background, generally results in a large deviation from the linear
 Hubble law, when there is a unique bang time.
Only  if $\Omega_0 \approx 0.001$, does a reasonable density contrast $A = 0.2$
produce an acceptable Hubble law.

As the above mentioned two pairs of initial conditions are interconnected,
one may start from the linear Hubble law as the velocity function, and
derive the required bang time function.
  This we have done, and conclude
that in this manner
 in the frame of the
LTB models it is thus possible to have a linear velocity -
distance relation when matter distribution is
fractal. Physically, this would mean that different spherical
 shells are
created at different moments given by the bang time function.

Thus the list of possible solutions of the Hubble-de Vaucouleurs
paradox (Baryshev et al. \cite{Baryshev}) in the frame of the
LTB description of fractality
now includes 1) a very low cosmic density, or 2) a dominating smooth
dark matter, or 3) non-simultaneus bang time.

\begin{table}
\begin{center}
\begin{tabular}{|c|c|c|c|c|}   \hline
$\Omega_0$ & $\xi_{{ZV}}$ & $R_{{ZV}} {(Mpc)}$ &
$\xi_{{fl}}$ & $R_{{fl}} {(Mpc)}$\\ \hline
$0.001$    & $0.001   $ & $6.3$          & $0.007$    & $33.6$        \\
$0.01$     & $0.012   $ & $61$           & $0.066$    & $330$         \\
$0.1$      & $0.1     $ & $530$          & $0.66$     & $3.3\,10^3$   \\
$0.99$     & $0.69    $ & $3.4\,10^3$    & $450$      & $2\,10^6$     \\
\hline
\end{tabular}
\end{center}
\caption{$A = 2$, $\delta(0) = 10^{6}$.}
\end{table}

\begin{table}
\begin{center}
\begin{tabular}{|c|c|c|c|c|}   \hline
$\Omega_0$ & $\xi_{{ZV}}$     & $R_{{ZV}} {(Mpc)}$ &
$\xi_{{fl}}$     & $R_{{fl}} {(Mpc)}$\\
\hline
$0.001$  & $1.2\,10^{-4}$ & $0.61$         & $6.7\,10^{-4}$ & $3.3$         \\
$0.01$   & $0.001$        & $6.1$          & $0.007$        & $33$          \\
$0.1$    & $0.01$         & $53$           & $0.06$         & $332$         \\
$0.99$   & $0.07$         & $344$    & $45$           & $2\,10^5$     \\
\hline
\end{tabular}
\end{center}
\caption{$A = 0.2$, $\delta(0) = 10^{5}$.}
\end{table}

\begin{table}
\begin{center}
\begin{tabular}{|c|c|c|c|c|}   \hline
$\Omega_0$ & $\xi_{{ZV}}$     & $R_{{ZV}} {(Mpc)}$ &
$\xi_{{fl}}$     & $R_{{fl}} {(Mpc)}$\\
\hline
0.001    & $9\,10^{-6}$   & $0.05$         & $6.3\,10^{-5}$ & $0.32$        \\
0.01     & $1.2\,10^{-4}$ & $0.6$          & $6.6\,10^{-4}$ & $3.3$         \\
0.1      & $0.001$        & $5.3$          & $0.007$        & $33$          \\
0.99     & $0.007$        & $34$           & $4.5$          & $2\,10^4$     \\
\hline
\end{tabular}
\end{center}
\caption{$A = 0.02$, $\delta(0) = 10^{4}$.}
\end{table}

\begin{table}
\begin{center}
\begin{tabular}{|c|c|c|c|c|}   \hline
$\Omega_0$   & $\xi_{{ZV}}$     & $R_{{ZV}} {(Mpc)}$ &
$\xi_{{fl}}$     & $R_{{fl}}{(Mpc)}$\\ \hline
$0.001$    & $complex$      & $complex$      & $3.4\,10^{-6}$ & $0.017$
\\
$0.01$     & $8.8\,10^{-6}$ & $0.04$         & $6.2\,10^{-5}$ & $0.31$        \\
$0.1$      & $10^{-4}$      & $0.5$          & $6.6\,10^{-4}$ & $3.3$         \\
$0.99$     & $6.8\,10^{-4}$ & $3.4$          & $0.45$         & $2\,10^3$     \\
\hline
\end{tabular}
\end{center}
\caption{$A = 0.002$, $\delta(0) = 10^{3}$.}
\end{table}

\begin{acknowledgements}

We thank the referee for valuable comments, C. Hellaby for kindly sending
us his paper, and M. Hanski for a useful discussion.
The work was supported by the Center for Cosmoparticle Physics
"Cosmion" (project "Cosmoparticle physics"), the Russian program
"Integration"(project
N.578), and the Academy of Finland
(project "Cosmology in the Local Galaxy Universe").

\end{acknowledgements}


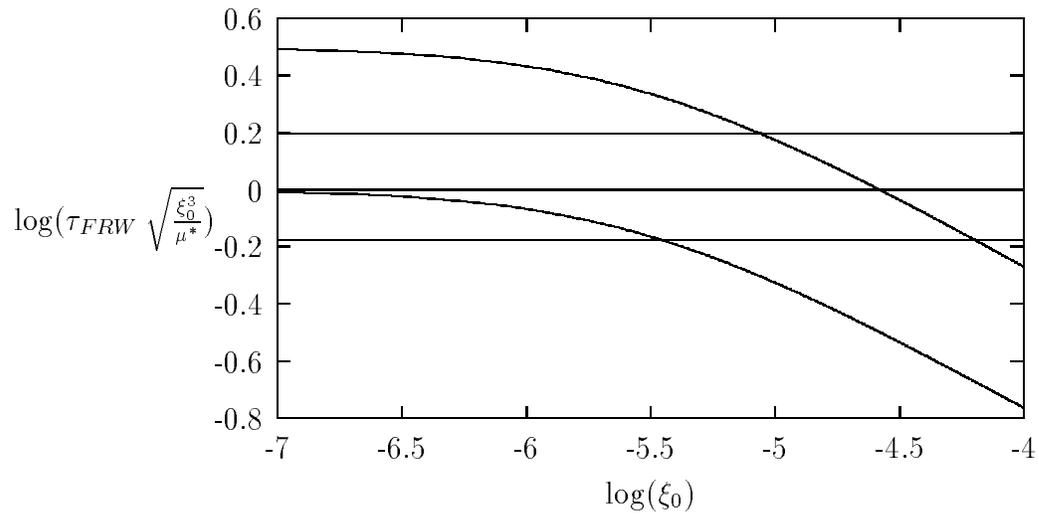
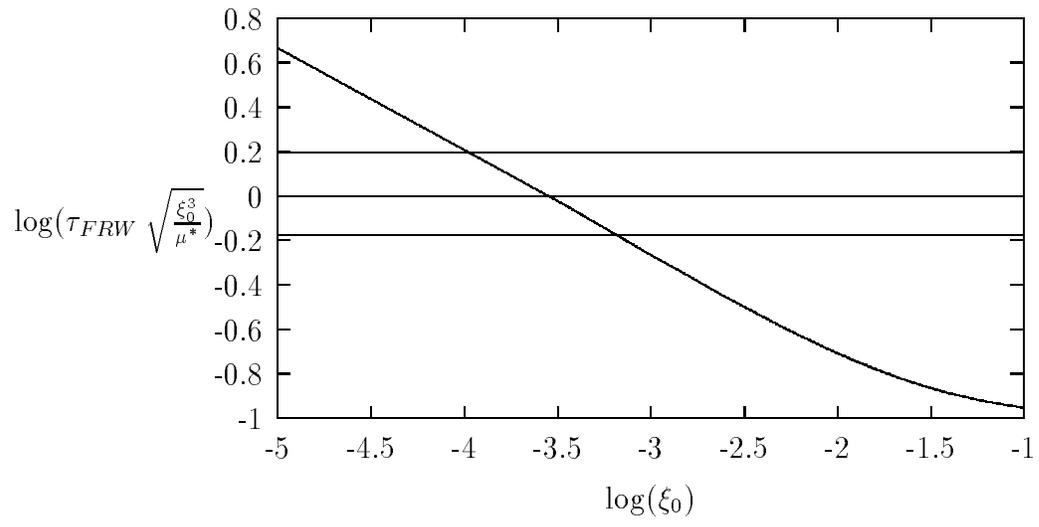
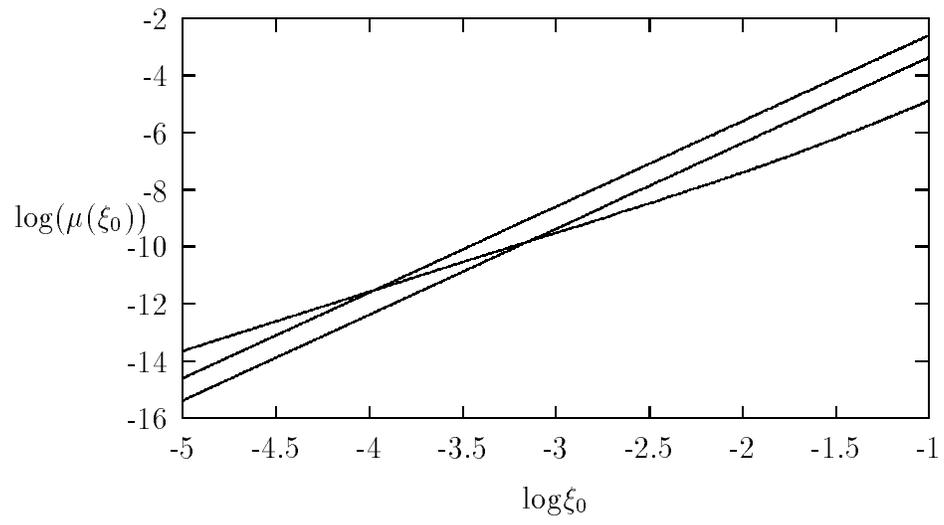
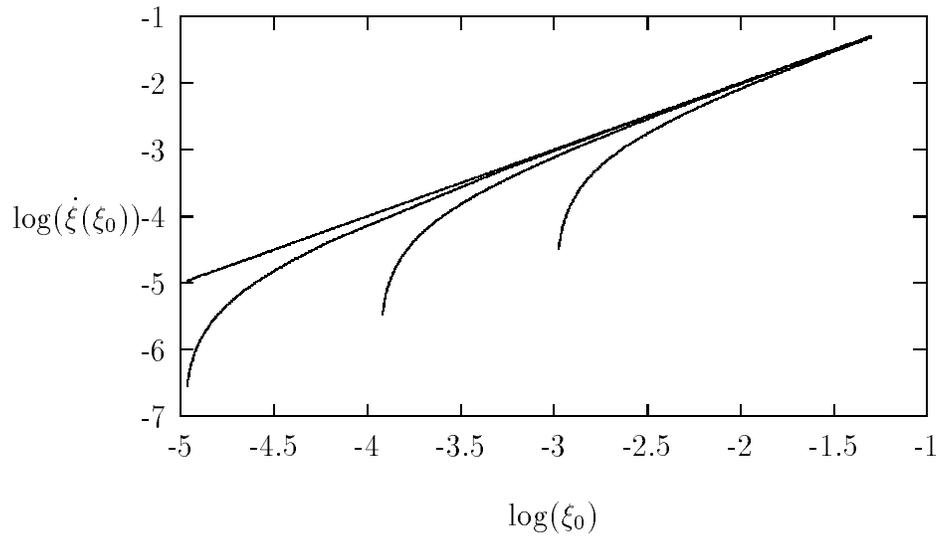
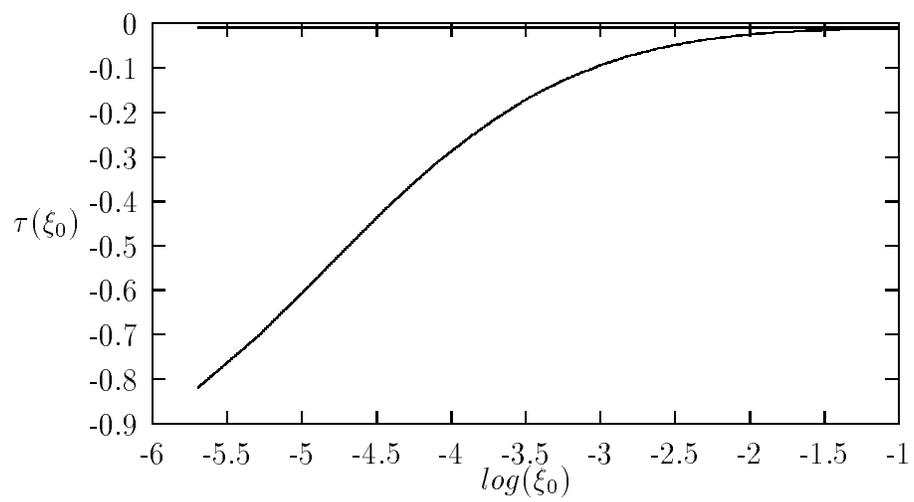
\end{document}